\documentclass[a4paper,fleqn]{cas-dc}

\usepackage[authoryear]{natbib}
\usepackage{amssymb}
\usepackage{amsmath}
\usepackage{float}
\usepackage{placeins}
\usepackage{xcolor}
\usepackage{comment}
\usepackage{listings}
\usepackage{ulem}
\usepackage{lineno}

\modulolinenumbers[5]

\AtBeginDocument{\renewcommand{\printorcid}{}}

\begin{document}
\shortauthors{Carrillo Navarro et~al.}

\title[mode=title]{A ready-to-deploy MLOps software platform for satellite and NEO detection at meter-class ground-based observatories}

\author[1,2,3]{Rafael Carrillo Navarro}
\cormark[1]
\ead{raficarrillo02@correo.ugr.es}

\author[4]{Pablo García-Martín}
\ead{pablo.garcia-martin@devinci.fr}

\author[3]{René Duffard}
\ead{duffard@iaa.es}

\author[1]{Juan Luis Orellana Montes de Oca}
\ead{jlorellana@roa.es}

\author[5]{Oscar Ortega Ríos}
\ead{oscar.ortegarios@alum.uca.es}

\affiliation[1]{
    organization={Real Instituto y Observatorio de la Armada (ROA)},
    addressline={Plaza de las Marinas s/n},
    city={San Fernando},
    postcode={11100},
    state={Cádiz},
    country={Spain}
}

\affiliation[2]{
    organization={Universidad de Granada, Departamento de Física Teórica y del Cosmos},
    city={Granada},
    postcode={18071},
    country={Spain}
}

\affiliation[3]{
    organization={Instituto de Astrofísica de Andalucía (IAA-CSIC)},
    addressline={Apdo. 3004},
    city={Granada},
    postcode={18080},
    country={Spain}
}

\affiliation[4]{
    organization={Safran, 171 Bd de Valmy},
    city={92700 Colombes},
    country={France}
}

\affiliation[5]{
    organization={Universidad de Cádiz, Departamento de Ingeniería Informática},
    city={Puerto Real},
    postcode={11519},
    country={Spain}
}

\cortext[1]{Corresponding author at: Universidad de Granada, Departamento de Física Teórica y del Cosmos, Granada, 18071, Spain.}

\begin{abstract}
Ground-based astronomical observations frequently contain streaks produced by artificial satellites, space debris, and potentially Near-Earth Objects (NEOs). While machine-learning models can reliably detect these features, their practical adoption in observatory operations is often limited by the lack of integrated tools for visual inspection, validation, workflow management, and structured data storage.

This paper presents the StreakMind Workbench, a framework that applies MLOps (Machine Learning Operations) practices to bridge research-oriented machine-learning pipelines with routine observatory operations. Rather than introducing new detection algorithms, the Workbench addresses a software-engineering challenge commonly encountered in astronomical computing: maintaining a single authoritative source of scientific processing code while providing astronomers with an operational environment for workflow execution and result inspection.

In this work, we show how this new platform achieves this goal through integration with the reference StreakMind AI model implementation described in \cite{Carrillo_26} ensuring that the scientific processing routines remain consistent with the underlying pipeline while avoiding unnecessary duplication of core processing logic. Implemented in Python using PyQt5, the Workbench supports the complete workflow from FITS ingestion to database storage, including inference execution, result inspection, database exploration, training management, and management of Minor Planet Center formatted observations.

Validation using 273 astronomical images acquired at La Sagra Observatory demonstrates the successful execution of end-to-end observational workflows, including inference, post-processing, database generation, and result inspection, while maintaining consistency with the underlying StreakMind original scientific code. The resulting platform facilitates the routine operational use of a research-oriented machine-learning pipeline in meter-class ground-based observatories and moderate-scale observational campaigns, supporting Space Situational Awareness and planetary defence activities.
\end{abstract}


\maketitle

\noindent\textit{Accepted manuscript of the article published in Astronomy and Computing
58 (2027) 101185.
DOI: \url{https://doi.org/10.1016/j.ascom.2026.101185}.
This version is the authors' LaTeX manuscript, not the Elsevier typeset PDF.}


\section{Introduction} \label{sec:Intro}

Ground-based astronomical observations play a dual strategic role in contemporary space science: they deliver critical data for planetary defence by detecting and characterising Near-Earth Objects (NEOs), while simultaneously supporting Space Situational Awareness (SSA) through the systematic monitoring of artificial satellites and orbital debris. Linear streaks produced by these fast-moving objects (whether natural asteroids or human-made spacecraft) have therefore become a scientifically valuable signal rather than an obstacle. Traditional streak-detection techniques, such as the Hough and Radon transforms or synthetic tracking, have long formed the backbone of moving-object searches. In recent years these classical methods have been powerfully augmented by automated pipelines and machine-learning approaches that bring greater sensitivity, scalability, and reproducibility to both planetary defence and SSA applications.

Recent advances illustrate the breadth of this progress. \cite{Stanescu_21} released the Umbrella open-source software suite, a modular end-to-end pipeline for asteroid detection, validation, identification, and reporting. Its principal innovation is a novel detection algorithm tailored specifically to faint trails, which has enabled successful near-real-time operations on ground-based facilities such as the Isaac Newton Telescope. Exploiting space-based archives, \cite{Woods_21} constructed a dedicated pipeline that processes full-frame images from the Transiting Exoplanet Survey Satellite (TESS) to automatically identify moving objects and submit astrometric measurements directly to the Minor Planet Center. On the machine-learning front, \cite{Duev_19} introduced DeepStreaks, a deep-learning framework utilising convolutional neural networks to efficiently identify fast-moving near-Earth objects in data from the Zwicky Transient Facility, significantly reducing the daily human workload required for streak identification from several hours to under 10 min. Extending these techniques to space-mission archives, \cite{HAH_3} applied machine-learning methods to detect and characterise asteroid streaks in Hubble Space Telescope observations, while a similar study did the same for archival data from the CHEOPS mission \citep{Garcia-Martin_26}, searching for images containing artificial satellites. Finally, hybrid methods that integrate machine learning with traditional image processing have been successfully deployed, as demonstrated by \cite{Stoppa_24}, whose ASTA pipeline pairs a U-Net model with a probabilistic Hough transform to robustly detect and delineate satellite trails in ground-based observations.

Building directly on this foundation, \cite{Carrillo_26} introduced StreakMind, a comprehensive AI-based end-to-end pipeline that employs a YOLO11 oriented bounding box (OBB) model trained on a hybrid real-synthetic dataset. The system automatically detects both satellite and NEO streaks in FITS frames, performs geometric refinement and inter-frame association, cross-identifies candidates against external satellite ephemerides, and stores all results in a normalised relational database. Evaluated on La Sagra Observatory data, StreakMind achieves 94\% precision and 97\% recall on an independent test set, demonstrating robust performance across a wide range of streak lengths, signal-to-noise ratios, and observing conditions.

While significant effort has been devoted to the development of machine-learning algorithms for astronomical image analysis, comparatively less attention has been given to the practical integration of these methods into operational observatory workflows. Bridging the gap between research-oriented pipelines and routine scientific operations remains an important challenge in astronomical computing, particularly when complex processing chains, database management, and result inspection must be combined within a single working environment. Building upon the StreakMind object detection model, the present work focuses on this complementary challenge: providing an operational platform capable of supporting its routine use in observatory settings. In practice, the adoption of StreakMind in observational workflows is greatly enhanced by interactive visualisation, structured result inspection, and streamlined workflow management. In this paper we present the StreakMind Workbench, a dedicated operational tool for integrating AI-based, existing object detection models (such as the StreakMind processing pipeline, \cite{Carrillo_26}), into observatory operations. The platform enables astronomers to overlay oriented bounding boxes, geometric reconstructions, and photometric elongation profiles directly onto the original FITS frames; to inspect cross-identification results and inter-frame track associations through a structured relational database; and to export all detection metadata for downstream analysis.

The present implementation is primarily intended for
meter-class ground-based observatories and moderate-scale
observational campaigns, rather than large survey facilities
employing complex multi-detector focal planes and high-throughput
processing architectures.

The contribution of this work is not the development of new streak-detection algorithms, which are described in \cite{Carrillo_26}, but the construction and validation of a dedicated environment that enables those algorithms to be executed, monitored, inspected, and managed within routine observatory workflows. The present paper therefore addresses software architecture, workflow integration, and operational validation rather than the underlying scientific processing methodology.

This work makes use of several open-source software packages, including Astropy \citep{astropy13,astropy18,astropy22}, Astroquery \citep{astroquery}, Ultralytics YOLO11 \citep{yolo11}, NumPy \citep{numpy}, Pandas \citep{pandas}, and Shapely \citep{shapely}. Additional tools include SQLite\footnote{\url{https://www.sqlite.org}} and PyQt5\footnote{\url{https://www.riverbankcomputing.com/software/pyqt/}}.

The remainder of this paper is organised as follows. Section~\ref{sec:architecture} describes the software architecture and design principles of the Workbench, its implementation, the containerisation and deployment strategy, the testing methodology, and the internal application architecture. Section~\ref{sec:functionality} summarises the processing workflows exposed through the platform. Section~\ref{sec:interface} presents the operational user environment and its principal workspaces. Section~\ref{sec:validation} reports the validation of the Workbench on observational data from La Sagra Observatory. Finally, we summarise the main findings, limitations, and future developments in Section~\ref{sec:conclusions}.


\section{Software Architecture} \label{sec:architecture}

\begin{figure*}[!tb]
\centering
\includegraphics[width=\textwidth]{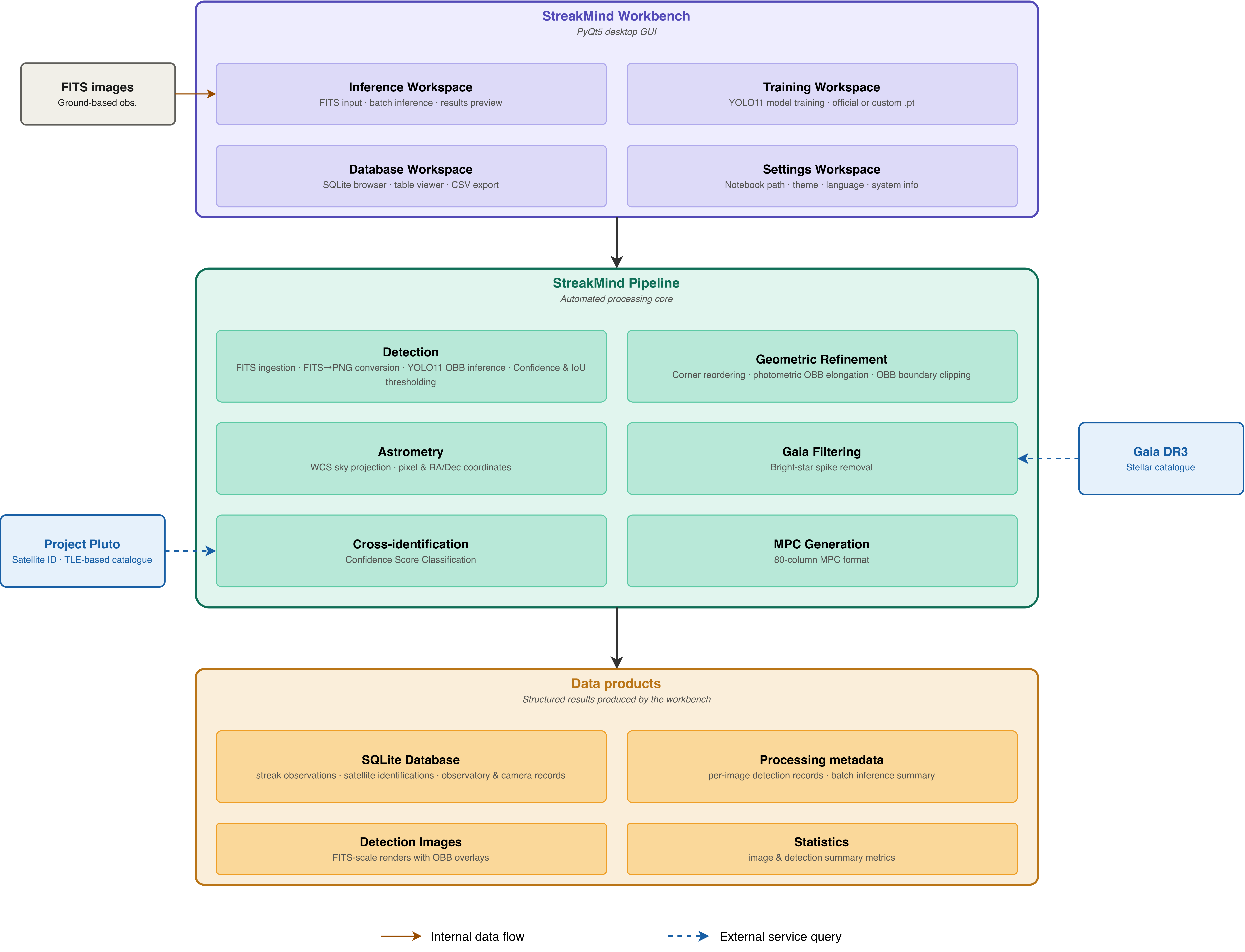}
\caption{Overview of the StreakMind Workbench architecture. The diagram illustrates the interaction between the operational environment provided by the Workbench, the underlying StreakMind processing pipeline, and the resulting data products generated during routine observational workflows.}
\label{fig:architecture}
\end{figure*}

Figure~\ref{fig:architecture} provides an overview of the StreakMind Workbench architecture and its interaction with the underlying StreakMind processing pipeline.

\subsection{Design principles}\label{sec:design_principles}
 
The StreakMind Workbench was designed around three guiding principles.

First, \textit{model agnosticism}: the Workbench does not embed any
detection logic of its own. All inference is performed by an object
detection model supplied by the user as a single \texttt{.pt} file
containing both the model architecture and the trained weights.
The Workbench interfaces directly with models implemented in the
Ultralytics YOLO framework, allowing users to switch between
different trained models without modifying the Workbench source code.
In the StreakMind pipeline described by \cite{Carrillo_26}, the
post-processing functions are loaded dynamically at runtime from the
reference pipeline notebook via the \texttt{jupyter-functions\allowbreak-exec}
package.

Second, \textit{instrument portability}: the Workbench
operates on standard FITS frames and their headers. All astrometric
information is extracted from the World Coordinate System (WCS)
solution embedded in each FITS file. This reliance on standard data products minimises observatory-specific configuration and facilitates portability to meter-class ground-based observatories
producing WCS-calibrated FITS images, within the instrumental and operational limitations discussed in Section~\ref{sec:conclusions}.

Third, \textit{output consistency}: every processing step in the StreakMind Workbench is designed to produce results consistent with those obtained when the model is used in a prototype or manual workflow (e.g., a Jupyter notebook). In our case, this is evaluated quantitatively using the StreakMind object detection pipeline described by \cite{Carrillo_26} in Section~\ref{sec:validation}.

Together, these principles enable the integration of inference, training, database exploration, and result visualisation within a unified operational environment, reducing the need for external tools during routine observatory use.
 
\subsection{Implementation and Pipeline Integration}

As illustrated in Figure~\ref{fig:architecture}, the Workbench is organised as an operational layer that interacts with the underlying StreakMind pipeline while exposing its functionality through a unified graphical environment.

The StreakMind Workbench is implemented in Python using the PyQt5
framework and follows a multi-threaded architecture. The main
application thread handles all Qt event processing and user
interactions, while computationally intensive operations including
FITS conversion, YOLO11 inference, post-processing, and database
writes, are executed in a dedicated worker. This separation ensures
that the graphical interface remains fully responsive during batch
processing of large image sets.

A key architectural decision was to avoid duplicating the pipeline logic between the StreakMind pipeline and the Workbench. The
\texttt{jupyter-functions\allowbreak-exec} package
parses the reference \texttt{.ipynb} file at startup, extracts function definitions from the notebook code cells, and makes them available
as callable Python objects. Every call to a pipeline function is
resolved against this parsed notebook and executed in the appropriate
Python context. The pipeline thus serves as the
authoritative reference implementation, while the Workbench
provides a graphical environment for executing and managing
the associated processing workflows.

\subsection{Setup and Configuration}

On first run, the interactive menu (\texttt{menu.sh}) presents a setup wizard that guides the user through the initial configuration. The user selects the execution mode, which is persisted in the \texttt{.env} file. The data directory path is requested when the platform or notebook is first launched, and is also stored in .env for subsequent sessions. The same menu provides options to toggle between modes at any time, to set up the required software environment (creating virtual environments on the host or building the Docker image), and to launch either the notebook pipeline or the software application.

All configurable parameters are centralised in a plain-text \texttt{.env} file located at the project root (see Table~\ref{tab:env_vars}).

\begin{table}[ht]
\centering
\caption{Environment variables exposed by \texttt{.env}.}
\label{tab:env_vars}
\begin{tabular}{ll}
\hline
\textbf{Variable} & \textbf{Purpose}\\
\hline
\texttt{RUNNING\_MODE} & Execution mode (\texttt{DOCKER} or \texttt{HOST})\\
\texttt{NOTEBOOK\_FILENAME} & Reference Jupyter notebook filename\\
\texttt{HOST\_DATA\_ROOT} & Data directory path on the host machine\\
\texttt{DATA\_ROOT} & Container mount point for data directory\\
\hline
\end{tabular}
\end{table}

\subsection{Multi-Environment Deployment}

The Workbench supports two complementary deployment strategies selected at runtime through the \texttt{RUNNING\_MODE} configuration variable. Container-based deployments require Docker Engine on the host system; GPU acceleration is supported on Linux hosts with the \texttt{nvidia-container\allowbreak-toolkit} runtime. Host-based deployments require Python 3.11 or later. The software has been tested with Python 3.11 and 3.12. On Windows, the software must be run inside a WSL2 environment with an Ubuntu distribution.

In \textit{Host mode}, the application runs directly on the host operating system using the native Python interpreter and Qt platform plugin --- \texttt{cocoa} on macOS and \texttt{xcb} on Linux. This mode is particularly suitable for development work and for deployments where Docker is unavailable or undesirable. The setup script scans for a compatible Python version (3.11 or 3.12) and warns the user if an untested version is detected, reducing the risk of environment-related failures.

In \textit{Docker mode}, the application runs inside a container with the software accessed through a VNC connection (\texttt{QT\_QPA\_PLATFORM="vnc:size=1920x1080:port=..."}). This mode provides a sandboxed, reproducible environment that isolates the application from variations in the host system's library stack, making it the recommended deployment method for observatory production use. Both modes are supported on macOS and Linux; on macOS, Docker does not support GPU acceleration via Metal Performance Shaders, so host mode is recommended when GPU inference is required on that platform, although Docker mode remains fully functional on CPU. The same GPU-acceleration detection and port-allocation mechanisms apply transparently to both the Workbench and the notebook server.

\subsection{Device Selection}

Device selection for model inference is handled automatically at runtime, prioritising CUDA-enabled GPUs, then Apple Metal Performance Shaders (MPS), and falling back to the CPU when no hardware accelerator is available. The user can override this automatic selection through the interface if required.

\subsection{Container Architecture}

For Docker-based deployments, the Workbench is conceived as a lightweight python 3.12 container image. The image includes the system libraries required by OpenCV (\texttt{libgl1}, \texttt{libglib2.0-0}) and PyQt5 (\texttt{libfontconfig1}, \texttt{libdbus-1-3}, \texttt{libqt5pdf5}) while keeping the image size to a minimum. The build process produces two image variants: \texttt{streakmind:0.0.1}, built for the native host architecture, and \texttt{streakmind-linux:0.0.1}, built with an explicit \texttt{-{}-platform=\allowbreak linux/amd64} target to ensure reproducible behaviour across different host architectures.

The application code is mounted as a Docker volume at \texttt{/app}, the data directory at \texttt{/data}, and user identity is synchronised through bind-mounted \texttt{.passwd} and \texttt{.group} files to prevent file-permission mismatches between the container and the host. When an NVIDIA GPU and the \texttt{nvidia-container-toolkit} runtime are available, GPU acceleration is enabled automatically through the \texttt{--gpus all} flag; otherwise, inference falls back to the CPU. An automatic port-allocation mechanism locates available ports for both the Jupyter notebook server and the VNC-based Workbench, avoiding conflicts with other services running on the host.

Two independent Python virtual environments are maintained: \texttt{.venv-gui} contains the PyQt5 framework and Graphical User Interface (GUI)-specific dependencies, while \texttt{.venv-nb} holds the Jupyter server and scientific libraries. This separation prevents dependency conflicts between the interactive analysis environment and the operational GUI, and reflects the dual nature of the system as both a research tool and a production deployment.

\subsection{Testing Strategy}
\label{sec:testing}

The testing methodology of the StreakMind Workbench is guided by the output consistency principle described in Section~\ref{sec:design_principles}: every processing step in the Workbench must produce results consistent with those of the reference StreakMind original pipeline. Testing therefore focuses on verifying that the graphical interface correctly invokes the same pipeline functions and produces identical outputs, rather than re-evaluating the scientific validity of the underlying detection algorithms.

\subsubsection{Test Dataset}

To support this approach, we use a representative test dataset of 20 FITS images drawn from the historical archive of La Sagra Observatory (MPC code L98) and centered on a single asteroid target. This dataset captures realistic observing conditions with varying exposure times, stellar backgrounds, and streak geometries. It serves as a lightweight validation suite for rapid regression testing during development; a larger dataset of 273 images is used for the full validation described in Section~\ref{sec:validation}.

\subsubsection{Inference Validation}

The inference workflow is validated by processing the test dataset through the Workbench's Inference workspace and comparing the resulting detections, Gaia-filtering outputs, satellite cross-identifications, and database records against those produced by the reference StreakMind pipeline under identical configuration parameters (confidence threshold 0.25, IoU threshold 0.45, input image size 640, maximum 300 detections per frame). Any discrepancy between the two outputs indicates a regression in the Workbench's integration layer rather than in the detection methodology itself.

\subsubsection{Training Validation}

The training workflow is validated by executing a lightweight training run (2 epochs, batch size 1, \texttt{yolo11n-obb} model) through the Training workspace and confirming that the generated configuration logs, optimisation curves, diagnostic plots, and exported model weights match those of a notebook-based training session with the same parameters. This test exercises the complete training pipeline --- dataset loading, model initialisation, forward and backward passes, logging, and model export --- without requiring the computational resources of a full training campaign.

\subsubsection{Cross-Platform Verification}

Because the test dataset and the accompanying \texttt{.env} configuration are platform-independent, the same validation procedures can be executed in both Docker and Host modes, providing a cross-platform safeguard against environment-specific regressions. Having a well-defined and stable test dataset ensures that the validation is fully reproducible across different machines and software versions.

\subsection{Application Architecture}
\label{sec:code_architecture}

Beyond the deployment and integration strategies discussed above, the internal organisation of the Workbench's source code follows a layered architecture designed to separate presentation, pipeline execution, and data persistence concerns. This section summarises the principal architectural patterns that support the operational workflows described in Section~\ref{sec:functionality}.

\subsubsection{Layer Organisation}

The application is structured in four logical layers. The presentation layer consists of the \texttt{MainWindow} container and five specialised tab widgets (Inference, Training, Database, Photometry, and Settings), each implemented as an independent \texttt{QWidget} subclass. The business logic layer is implemented through background worker threads that orchestrate the computationally intensive stages of each workflow, leaving the main application thread free to process user interactions. The integration layer bridges the Workbench to the reference StreakMind pipeline described by \cite{Carrillo_26} through a dedicated notebook function wrapper that parses the reference \texttt{.ipynb} file at startup and exposes its function definitions as callable Python objects. Finally, the data access layer uses direct SQLite connections without an object-relational mapper, with a thin utility module (\texttt{db\_utils.py}) handling all database write operations to the schema described in Section~\ref{sec:functionality}.

\subsubsection{Configuration Management}

The Workbench employs a dual configuration strategy. Environment-dependent path variables such as the data directory, running mode, and reference notebook filename are managed through a plain-text \texttt{.env} file consumed at startup by two Python modules: \texttt{config\_paths.py}, which defines the complete data directory tree and validates that the configured data root exists, and \texttt{name\_cuaderno\_jupyter.py}, which resolves the path to the reference notebook. In contrast, user interface preferences including theme selection, font size, interface language, and last-used file paths are persisted independently through Qt's \texttt{QSettings} mechanism, ensuring that runtime configuration changes within the software do not affect the underlying pipeline paths.

\subsubsection{Inter-Component Communication}

Communication between the Workbench's components relies on Qt's signal--slot mechanism, which provides type-safe, decoupled message passing. Worker threads emit progress signals to update the interface during long-running tasks, while the completion of an inference batch automatically triggers downstream tabs (Database and Photometry) to refresh their content with the newly written results. Global interface changes, such as a language or theme switch, are propagated through a full user interface rebuild orchestrated by the \texttt{MainWindow}, ensuring that all widgets are instantiated with the updated configuration without requiring individual restart.

\subsubsection{Multi-Threading Strategy}

Two distinct threading models are employed depending on the nature of the workload. For inference and photometry processing, which interact with the Qt event loop through progress signals and preview updates, standard \texttt{QThread} subclasses are used; the worker's \texttt{run()} method executes the pipeline sequentially while emitting state updates to the main thread. For model training, however, a \texttt{QThread} wrapper launches a separate \texttt{multiprocessing.Process}, as training can occupy the device for extended periods and must be terminable from the interface regardless of the compute backend in use. The global process start method is set to \texttt{"spawn"} to ensure compatibility with PyQt5 across all supported platforms.

\subsubsection{Internationalisation and Theming}

The Workbench supports dynamic switching between English and Spanish through a dedicated \texttt{lang} module that exposes a \texttt{tr()} function used by every widget to resolve interface strings. Language changes are applied by destroying and recreating all tabs, which forces a complete re-read of the translated strings without requiring an application restart. The visual appearance is controlled by a QSS-based theming system offering eight prebuilt colour schemes together with support for a user-defined accent colour. A global font-size slider adjusts the text size across the entire application by rewriting the active stylesheet.

\subsection{Software Engineering Challenges}
\label{sec:engineering_challenges}

Several non-trivial software-engineering challenges arise when embedding a research-oriented machine-learning pipeline within an interactive desktop application. This section summarises the most significant of these and the strategies adopted to address them.

\textbf{Pipeline--Workbench code duplication.} The core detection, post-processing, and cross-identification logic originates in the reference Jupyter notebook described by \cite{Carrillo_26}. Duplicating this code in the Workbench would create a maintenance burden and risk divergence between the research pipeline and the operational tool. The solution adopted is a runtime bridge: the \texttt{jupyter-functions\allowbreak-exec} package parses the \texttt{.ipynb} file at startup, extracts function definitions from code cells, and exposes them as callable Python objects. Module-level imports within the notebook that reference project-internal paths are satisfied through the injection of dummy placeholder modules into \texttt{sys.modules}, preventing import errors while keeping the notebook unchanged. This architecture ensures that the notebook remains the authoritative implementation and that any update to the pipeline is immediately available in the Workbench without additional effort.

\textbf{Training termination across compute backends.} The YOLO training loop occupies the selected compute device for extended periods and must be interruptible from the graphical interface. The standard approach of subclassing \texttt{QThread} and calling \texttt{terminate()} is unreliable on GPU backends such as CUDA and Apple Metal Performance Shaders, where thread-level termination cannot release device resources cleanly. The Workbench addresses this by launching training inside a separate \texttt{multiprocessing.Process}, created within a \texttt{QThread} wrapper. The process boundary guarantees that \texttt{process.terminate()} releases all device allocations regardless of the backend, while the wrapper thread relays status updates to the interface through standard Qt signals. The global process start method is set to \texttt{"spawn"} to avoid deadlocks between \texttt{fork}-based process creation and the Qt event loop.

\textbf{Gaia catalogue query efficiency.} Bright-star spike filtering requires querying the Gaia DR3 catalogue for sources brighter than $G_\mathrm{RP} = 6.0$ in each observed field. Issuing a TAP query per image would be prohibitively slow for large batches and would risk degrading the Gaia TAP service under rapid repeated requests. The solution groups images by observing session and issues one query per session, covering the aggregate sky region of all frames acquired during that night. A configurable retry mechanism with a fixed 5-second delay between attempts handles transient network or service failures transparently.

\textbf{Trail extent underestimation.} Oriented bounding box regressors in detection models are optimised for compact objects and systematically underestimate the true extent of linear features that span a significant fraction of the field. The Workbench corrects this in a post-processing step: for each surviving detection, a photometric profile is extracted along the trail's major axis and the bounding box endpoints are extended until the signal falls below a locally estimated background threshold. This refinement operates on pixel data rather than model predictions and ensures that the derived trail length, endpoints, and position angle reflect the actual extent of the detected streak.

\textbf{Internationalisation without restart.} Conventional internationalisation approaches require loading all translated strings at application start, making runtime language switching difficult to implement correctly. The Workbench adopts a different strategy: interface strings are resolved lazily through a \texttt{tr()} function, and a language change triggers a complete destruction and recreation of all tab widgets via the \texttt{MainWindow.rebuild\_ui()} method. Each widget is re-instantiated with fresh string lookups, guaranteeing that every visible element reflects the new language without requiring an application restart or per-widget update logic.


\section{Processing Workflows}
\label{sec:functionality}

\subsection{Image Ingestion}\label{image_ingestion}

The StreakMind Workbench accepts individual FITS files or entire
directories as input, supporting both single-frame inspection and
large batch campaigns. Before inference, each frame is
automatically converted to a normalised image following the same
procedure as the StreakMind pipeline described by
\cite{Carrillo_26}. A ZScale contrast stretch, as
implemented in \texttt{astropy.visualization\allowbreak-exec},
is applied to enhance the visibility of faint features while
avoiding saturation from bright sources, an approach widely used
in astronomical image display. 
The original FITS file is always retained as the reference for all astrometric
computations; the converted image is a temporary working copy used
exclusively during the detection step. Instrumental metadata
including observation date, exposure time, telescope identifier,
and observatory code are extracted from the FITS header at
ingestion and stored alongside the detection results, providing
full traceability from raw frame to final output.

\subsection{Streak Detection}

Detection is performed by an object detection model compatible
with the Ultralytics open framework, supplied by the user as a
\texttt{.pt} file. This design allows any Ultralytics-compatible
model to be used without modifying the Workbench source code.
In the present implementation this corresponds to a YOLO11
Oriented Bounding Box model trained on a hybrid real and synthetic dataset
of astronomical streaks, as described in \cite{Carrillo_26}.
All inference parameters, including confidence threshold, IoU
threshold, input image size, maximum number of detections per
frame, and compute device, are fully configurable through the
interface and are set to sensible defaults. Device selection is
automatic, prioritising GPU acceleration when available, but can
be overridden manually, making the Workbench deployable on both
dedicated GPU workstations and standard laptop hardware. 
The downstream processing pipeline, however, is
designed specifically for artificial satellite and space debris
trails, following the StreakMind pipeline described by
\cite{Carrillo_26}. Extension to NEO detection is discussed in
Section~\ref{sec:conclusions}.

\subsection{Geometric Reconstruction and Astrometry}

Each surviving detection undergoes geometric reconstruction to
derive the trail endpoints and centroid, both in pixel coordinates
and in equatorial coordinates (RA, Dec), together with the
position angle of the trail on the sky. Each trail is also
classified as complete or incomplete depending on whether both
endpoints are clearly defined within the frame. Celestial
coordinates are computed using the astrometric solution embedded
in the FITS header. Accurate astrometry is essential for
downstream cross-identification against object catalogues:
small coordinate errors can lead to false matches or missed
identifications, particularly for fast-moving objects observed
at high angular rates. All geometric and astrometric quantities
are stored in the output database.

\subsection{Bright-Star Spike Filtering and Photometric Trail Refinement}\label{spike_filter}

Diffraction spikes from bright stars are a primary source of false
positives in streak detection, as their elongated shape closely
mimics that of a genuine trail. After batch processing, the
Workbench queries the Gaia DR3 catalogue \citep{Gaia_23} to
identify bright stars in the observed field and automatically
flags detections that lie within $450^{\prime\prime}$ of any
source brighter than $G_\mathrm{RP} = 6.0$. Queries are issued once per
observing session rather than per image, minimising network load
and avoiding degradation of the Gaia TAP service under rapid
consecutive requests. Flagged detections are excluded from all
subsequent steps. This automated filtering step removes the need
for manual inspection of spike-contaminated detections, which can
represent a significant fraction of the raw output in fields
containing bright stars, as demonstrated in
Section~\ref{sec:validation}. A retry mechanism handles transient
service failures transparently, ensuring that a temporary network
outage does not interrupt a running batch.

For each detection that survives the filter, a photometric profile
along the trail axis is extracted to refine the geometric
reconstruction. As described in \cite{Carrillo_26}, YOLO-based
regressors frequently underestimate the true extent of linear
streaks, as they are not optimised for objects that span a
significant fraction of the field of view. To address this, the
endpoints of the oriented bounding box are adjusted to the actual
extent of the trail by growing the box along its major axis until
the signal drops below a locally estimated background threshold,
ensuring that the derived trail length and endpoints accurately
reflect the true extent of the detected streak rather than the
initial model output.

\subsection{Satellite Cross-Identification and MPC Reporting}

For each detection that survives the filtering step, the Workbench automatically generates an observation line in Minor Planet Center (MPC) format\footnote{\url{https://www.minorplanetcenter.net}}, the standard reporting format accepted by the international astrometric community for both natural and
artificial moving objects. The MPC line encodes the observation
epoch, the trail centroid position in equatorial coordinates, an
estimated magnitude, and the observatory code.
The observation epoch corresponds to the start of the exposure, as recorded in the FITS header.
The observatory
code is read from the FITS header when available, or can be
entered manually in the interface. All MPC lines from a batch are submitted to the Project Pluto
satellite identification service \citep{ProjectPluto2025},
which cross-matches them with the current catalogue. The
identification results, including satellite name, angular
offset from the predicted position, and a confidence score,
are stored in the output database alongside the detection
record. This enables rapid assessment of whether a detected
trail corresponds to a known catalogued object or represents
an uncatalogued or unidentified track.

\subsection{Database Storage and Export}

All pipeline outputs are stored in a structured SQLite relational
database that persists across sessions and can be shared between
observers. The schema organises results across seven tables covering
detections, processed images, satellite identifications,
instrumental and observatory metadata, and per-session image
statistics. Each detection record is uniquely identified by its
MPC observation line, ensuring that results from multiple
processing runs can be combined without duplication. This
design prevents repeated processing of the same observations
from generating duplicate database entries while preserving
traceability between detections, images, and associated metadata. The database
is designed to support both individual observatory operations and
collaborative multi-site campaigns, where observations from
different facilities can be merged into a common archive. Any
table can be exported to JSON for downstream analysis, submission
to external catalogues, or integration with observatory data
management systems.


\section{Operational Workspaces}
\label{sec:interface}

The StreakMind Workbench provides four operational workspaces
corresponding to the principal stages of the processing workflow:
inference, training, database exploration, and configuration
management. Together, these workspaces provide access to the core
capabilities of the platform within a unified operational environment.

\subsection{Inference Workspace}

\begin{figure*}[!tb]
\centering
\includegraphics[width=\textwidth]{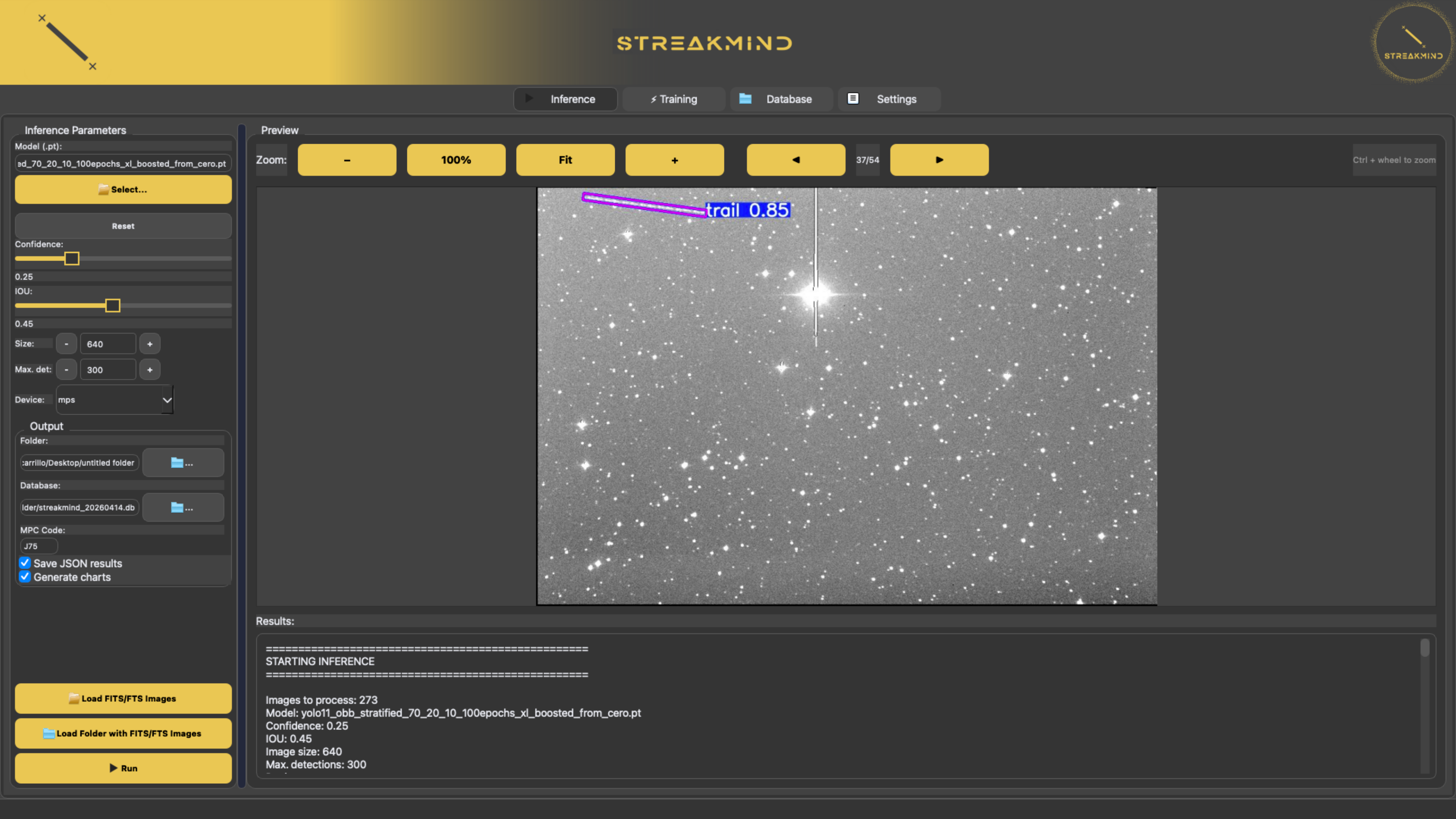}
\caption{Overview of the \textit{Inference} workspace of the StreakMind
Workbench during the inference run performed for the validation
dataset from La Sagra Observatory. The interface is organised into
a left-hand control panel for model selection, inference parameters,
output configuration, and execution controls; a central preview
region for image display, graphical overlays, zoom controls and
navigation between processed images; and a lower results panel
reporting the progress and status of the processing run in real time.}
\label{fig:inference_tab}
\end{figure*}

The \textit{Inference} workspace is the primary operational environment
of the StreakMind Workbench (Figure~\ref{fig:inference_tab}) and
provides access to the full workflow required to run streak detection
on individual FITS frames or on large batches of images. The layout
is divided into a left-hand control panel and a right-hand workspace
composed of an image preview region and a results panel.

The control panel exposes all parameters required for the detection
step. These include the path to the model weights file, confidence
threshold, Intersection over Union (IoU) threshold, input image size,
maximum number of detections per frame, and compute device selection.
The confidence threshold defines the minimum confidence required for
a predicted object to be retained; lower values increase sensitivity
but may introduce false positives. The IoU threshold measures the
overlap between two bounding boxes, defined as the area of their
intersection divided by the area of their union. The input image size
corresponds to the spatial resolution used by the detection model:
the input image is resized to a square of size $N \times N$ pixels,
where $N$ is the selected value, with padding applied to preserve the
aspect ratio when the original frame is not square; larger values may
improve detection of small objects but increase computational cost.
Compute device selection determines the hardware backend used during
inference; the Workbench dynamically exposes only the devices
supported by the host system, including CPU, CUDA-enabled GPUs, and
Apple Metal Performance Shaders (MPS), with the most suitable
available device preselected automatically, although the user may
override this choice. A reset option restores all parameters to
their default values. Output configuration is handled
in the same panel, where the user specifies the destination directory
for the processing results, the SQLite database in which detections
will be stored, and the MPC observatory code. The selected database
can be either newly created or an existing one, allowing results from
multiple processing sessions to accumulate in a single persistent
archive. Optional controls allow the generation of JSON summaries and
statistical plots summarising the batch results.

As described in Section~\ref{sec:functionality}, images are ingested
from individual FITS files or entire directories. Within the
\textit{Inference} workspace, the selected data can be visually
inspected in the preview region before launching the detection
pipeline. A dedicated conversion control performs the
FITS-to-PNG transformation following the procedure described in
\cite{Carrillo_26}, and is always executed prior to inference to
provide the image format required by the detection model. Once the
inference parameters and output configuration have been defined, the
detection process can be launched directly from the interface.

During execution, the preview region displays the currently processed
image together with the graphical overlays generated by the pipeline,
including oriented bounding boxes and the reconstructed streak
geometry. These overlays correspond to the
pre-filtering stage of the workflow, before the Gaia-based
bright-star spike filtering described in
Section~\ref{spike_filter}, since the filtering is applied
collectively over grouped sky regions at the end of the batch
rather than on a per-image basis. At the same time, the results
panel provides a chronological textual log of the processing
session, reporting the progress of the inference workflow and
summarising the outcome of each processing step. When the batch
completes, the panel reports the overall processing summary,
including the results of the Gaia-based bright-star filtering and
the satellite cross-identification stage.

The preview region also provides interactive visualisation tools,
including zoom, and fit-to-window controls that allow the user to
inspect detections at different spatial scales. Once the inference
run has completed, the same panel can be used to navigate through
the processed images of the batch, allowing rapid visual inspection
of the detections associated with each frame without leaving the
main workspace.

The inference workflow produces a structured set of
output products. These include image visualisations of the
detections at each stage of the processing pipeline, from the
initial model output through geometric and photometric
post-processing. Statistical summaries describing the overall
behaviour of the processed dataset are also generated,
covering distributions such as the number of detections per
image, the fraction of images containing streaks, and the
occurrence of multiple detections within the same frame.
All detection records are exported in JSON format, both at the
individual-image level and as a global batch summary, and all
results are stored in the selected SQLite database, which
serves as the persistent archive of the inference session.

\subsection{Training Workspace}

The \textit{Training} workspace provides an interface for training
the detection model on new datasets, allowing users to configure and
launch the training process directly from the Workbench without
requiring command-line interaction.


The control panel exposes the main parameters governing the
training procedure. The dataset is specified through a YAML
configuration file, while the user may select either a standard
YOLO11 base model or a custom pretrained model as the starting
point. Configurable parameters include the number of training
epochs, batch size, input image size, number of data-loader
workers, and compute device. The output directory and experiment
name can also be defined prior to launching the training run.

During execution, a real-time output console streams the training
log generated by the underlying framework, while a progress bar
indicates the overall completion of the training process.
Training can be interrupted at any time from the interface
if required.

The training procedure automatically generates a dedicated output
directory containing the stored training configuration, training
metrics and logs, diagnostic plots describing the optimisation
behaviour (e.g. precision--recall curves, F1-score curves, and
confusion matrices), example batches of labelled images and model
predictions, and the resulting trained model weights. This
structure facilitates the inspection, comparison, and
reproducibility of training experiments across different
datasets and observing conditions.

\subsection{Database Workspace}

The \textit{Database} workspace provides an interactive viewer for
the SQLite output database generated during inference, allowing users
to browse and inspect the stored processing results directly within
the Workbench environment.


The upper section of the interface contains a table selector that
allows the user to navigate between the different tables of the
database, whose relational structure follows the schema described
in \cite{Carrillo_26}. The main workspace displays the contents
of the selected table in a scrollable grid view, where each row
corresponds to a database record and each column represents one
of the fields defined in the table schema. Selecting a row
automatically populates a preview panel on the right-hand side
of the interface, displaying all column values of the selected
entry in an expanded format for easy inspection.

Several management controls are also available. The active
database file can be replaced at runtime to inspect results from
different processing sessions, and the view can be refreshed
after new data are written during inference. The selected table
can be exported to JSON format for external analysis. A reset
option allows the user to clear the contents of the current
database after confirmation, with an optional step to also
remove the associated output directory.

\subsection{Settings Workspace}

The \textit{Settings} workspace centralises the persistent
configuration parameters of the StreakMind Workbench, allowing the user
to adjust interface preferences, define the reference notebook
used by the pipeline, and inspect the runtime software
environment.

A font-size slider allows the text size to be adjusted
dynamically across the application. The \textit{Notebook} panel
allows the user to specify the path to the reference Jupyter
notebook containing the core pipeline functions used by the
Workbench, keeping the graphical interface decoupled from the
scientific processing code and allowing the pipeline
implementation to be updated without modifying the software.

The \textit{System information} panel displays the current
runtime environment, including the installed versions of Python,
PyTorch, and Ultralytics, together with the available compute
backend. The \textit{Appearance} panel allows the visual theme
to be configured, while the \textit{Language} panel provides a
selector for the interface language. At present, the Workbench
supports English and Spanish, and language changes are applied
dynamically within the running application.


\section{Validation and Real-World Application}\label{sec:validation}

The purpose of this section is not to evaluate the detection performance of the underlying machine-learning methodology, which is analysed in detail by \cite{Carrillo_26}. Instead, the objective is to validate the operational execution of the complete StreakMind workflow within the Workbench environment, including inference, post-processing, database generation, satellite identification, and model-training tasks.

To this end, a representative dataset obtained at
La Sagra Observatory (MPC code L98) was used as a
real-world test case. The dataset consists of 273
astronomical images acquired during routine asteroid
observations. Specifically, it includes 99 frames
corresponding to the observation of asteroid
(0393) Lampetia on the night of 1 August 2019,
80 frames obtained during the same night for the
observation of asteroid (0472) Roma, and 94 frames
acquired during the observation of asteroid (319) Leona on 13 February 2024. All images obtained
during August 2019 were taken with an exposure
time of 60 s, whereas the frames acquired
in February 2024 have an exposure time of
100 s. Although these images were originally
obtained for asteroid photometry, many of them
contain moving-object streaks crossing the field of
view. Such data therefore provide a realistic
scenario for validating the operational behaviour
of the StreakMind Workbench under typical observing
conditions, including stellar backgrounds, varying
streak geometries, and multiple detections within
the same frame.

The validation focused on both the inference and training
workflows implemented in the StreakMind Workbench.
For the inference validation, the dataset was processed
using the \textit{Inference} workspace of the Workbench with
the same detection model and configuration parameters
employed in the original StreakMind pipeline. These
parameters included a confidence threshold of 0.25,
an Intersection over Union threshold of 0.45, an input
image size of 640 pixels, and a maximum of 300 detections
per frame. The inference was executed on the MPS device
backend and the MPC observatory code was set to L98.

The validation verified the successful execution of the complete inference workflow, including image ingestion, streak detection, geometric reconstruction, Gaia-based filtering, satellite cross-identification, MPC observation generation, and database population. The resulting outputs were compared against those produced by the reference StreakMind implementation described in \cite{Carrillo_26}, confirming consistent behaviour across all stages of the workflow.

The training workflow was validated in a similar way
using the \textit{Training} workspace of the Workbench. A
lightweight test training run was launched through the
graphical interface using the same dataset configuration
file employed by the StreakMind pipeline. The objective of this experiment was not to optimise model performance
but to verify the correct execution of the training workflow
within the Workbench environment.

To minimise computational requirements while exercising
the full training pipeline, a reduced configuration was
used consisting of 2 training epochs, a batch size of 1,
an input image size of 640 pixels, and zero data-loader
workers. The training process was executed on the MPS backend. The initial model was the
pretrained \texttt{yolo11n-obb\allowbreak-exec} architecture, corresponding
to the smallest variant of the YOLO11 oriented bounding
box models in terms of parameter count.

The training logs, optimisation curves, evaluation plots,
and resulting model weights generated through the
Workbench were compared with those produced by executing
the same training procedure directly with the StreakMind
pipeline. The experiment confirmed the correct execution
of the training workflow through the Workbench environment,
including configuration handling, training launch,
monitoring, output generation, and model export.
Minor variations may arise across software versions and
hardware platforms due to differences in numerical
implementations, as well as from the non-deterministic
nature of remote catalogue queries, which may return
slightly different source lists across executions;
however, these effects are negligible and do not affect
the overall results.

Figure~\ref{fig:inference_tab} illustrates a representative example
of the StreakMind Workbench operation during the validation tests.

These validation tests demonstrate that the StreakMind Workbench supports the end-to-end execution of the complete StreakMind workflow within a unified operational environment. The system successfully integrates inference, post-processing, database generation, satellite identification, training management, and result inspection while maintaining consistency with the underlying methodology described in \cite{Carrillo_26}.

To illustrate the practical use of the system in a real
observational scenario, the preview panel shown in
Fig.~\ref{fig:inference_tab} is used as a representative
example from the validation dataset. In this case, a
satellite streak crossing the field of view is detected
by the neural network and displayed together with the
graphical overlays produced by the pipeline, including the
oriented bounding box predicted by the model (blue) and
the reconstructed streak geometry (pink). The corresponding
detection parameters are subsequently stored in the database
together with the associated observation metadata, enabling
further analysis and cross-identification with external
satellite catalogues.

In this particular case, the detected streak corresponds
to an object that appears in a sequence of five consecutive
images in the dataset, allowing the reconstruction of a
consistent track across multiple frames. The derived astrometric and geometric parameters are stored in the \textit{observations} table of the output database, as illustrated in Table~\ref{tab:example_observation}, which shows the entry associated with one of these detections,
including the reconstructed central coordinates of the
streak, the endpoints defining the oriented bounding box,
the position angle of the track, the MPC-formatted
observation line, and other associated observational
metadata.

\begin{table}[t]
\centering
\caption{Example entry from the \textit{observations} table
generated by the StreakMind Workbench during the validation
experiment.}
\label{tab:example_observation}
\vspace{6pt}
\begin{tabular}{ll}
\hline
Parameter & Value \\
\hline
Date & 2019-08-01 \\
MPC Date & 2019 08 02.04451 \\
MPC Code & L98 \\
Trail Type & Complete \\
Track ID & MLD0036 \\
RA Central & 23 22 47.93 \\
DEC Central & -14 53 45.6 \\
RA Mark1 & 23 23 19.15 \\
DEC Mark1 & -14 52 42.8 \\
RA Mark2 & 23 22 16.70 \\
DEC Mark2 & -14 54 48.3 \\
Pixel Mark1 & (348.47, 683.63) \\
Pixel Central & (458.44, 670.11) \\
Pixel Mark2 & (568.40, 656.60) \\
Position Angle & 97.01 deg \\
Observer & Nicolas Morales \\
Analyst & Rafael Carrillo \\
\hline
\end{tabular}
\end{table}

Using these reconstructed parameters, the Workbench
identifies the object as the satellite \textit{INTELSAT 605}
(NORAD ID 21653, COSPAR designation 1991-055 A). The
identification is obtained by comparing the reconstructed
streak position with predicted satellite ephemerides
using the Project Pluto tools, as described in
\cite{Carrillo_26}. For this object, the angular offset
between the predicted and observed positions is
$0.0285^\circ$, corresponding to an identification
probability of $99.55\%$. This example illustrates
how the Workbench enables the detection, characterisation,
and identification of moving-object streaks in astronomical images while storing the resulting measurements in a
structured database suitable for further analysis and
catalogue comparison.


\section{Conclusions, limitations, and future work}\label{sec:conclusions}

The StreakMind Workbench provides an operational platform that applies MLOps practices to integrate existing AI-based object detection models into routine activities at meter-class ground-based observatories. Validation using an existing AI model and a representative dataset from La Sagra Observatory demonstrated the successful integration of image ingestion, streak detection, post-processing, database generation, satellite cross-identification, result inspection, and model-training workflows within a unified platform.

By combining automated processing, interactive visualisation, and structured data management, the Workbench facilitates the practical exploitation of astronomical datasets or real-time analysis of images containing satellite or other moving-object streaks. This capability enables observations originally acquired for different scientific objectives to be reused for the detection, characterisation, and identification of artificial objects, contributing to Space Situational Awareness (SSA) and planetary defence activities.

The current implementation nevertheless presents several limitations that define its present operational scope:

\begin{itemize}

\item The workflow has not been designed to reconstruct individual streaks spanning multiple detectors, as may occur in large focal-plane mosaic cameras.

\item Satellite cross-identification depends on the accuracy
of the available orbital information. Although the confidence-scoring
scheme developed in the original StreakMind pipeline helps rank
candidate identifications based on their angular offsets, multiple
plausible candidates may still remain when their predicted positions
are similarly close to the observed streak. Brightness information is
not currently used as an additional criterion to distinguish between
such candidates.

\item The Gaia-based bright-star filtering stage has only been validated under the observational conditions represented in the
current dataset. Its behaviour in substantially deeper fields or in
fields with markedly different Gaia DR3 source densities has not yet
been characterised. Calibrated trail photometry is not provided by
the current implementation.

\item Astrometric coordinates are derived from the WCS solution provided in the FITS headers, and the Workbench does not compute an independent local plate solution. Consequently, inaccuracies in the input WCS may propagate into the reconstructed astrometry and subsequent cross-identification.

\item The Workbench generates MPC-formatted astrometric records, while the acceptance and submission requirements for isolated streak measurements or short tracklets are determined by the MPC and remain external to the platform.

\item Although the Workbench allows different Ultralytics-compatible models to be supplied by the user, the reference StreakMind model employed in the present validation has been trained and characterised using data from La Sagra Observatory and has not yet been independently evaluated on datasets from multiple observatories. Consequently, no multi-observatory confusion-matrix analysis is presented in this work.

\item The current integration strategy deliberately retains the reference Jupyter notebook as the authoritative implementation of the scientific processing functions. Although this avoids duplication and divergence between the research pipeline and the Workbench, it may be less suitable for highly automated or large-scale processing environments than a fully packaged processing backend.

\end{itemize}

The current design naturally supports further extensions
towards enhanced scientific exploitation of the detected
streaks, and ongoing developments are focused on expanding
these capabilities. In particular, the extraction of
calibrated photometric measurements along the reconstructed trails is being implemented to enable a more detailed characterisation of detected objects and to provide an additional criterion for resolving ambiguous satellite identifications. Similarly, the identification framework is being extended to include associations with small Solar System bodies, allowing detected streaks to be linked to candidate asteroid trajectories through probabilistic matching schemes
analogous to those used for satellite identification.

Further developments will also address some of the
limitations identified above. In particular, multi-observatory
validation and, where necessary, fine-tuning of the reference
detection model will be investigated to characterise its
performance under different instrumental and observational
conditions. Support for multi-detector configurations and
optional verification or refinement of the input astrometric
solution will also be explored. Any future implementation of
direct MPC submission will remain aligned with the applicable
MPC reporting requirements.

From an operational perspective, continued development of the
Workbench is focused on extending multi-language support and improving
computational performance, enhancing both usability and scalability for moderate-scale observational campaigns.

For deployments requiring a higher degree of automation
and scalability, future development may also consider
refactoring the core scientific processing functions into a
versioned Python package while preserving the current
single-authoritative-source principle.


\section*{CRediT authorship contribution statement}
Rafael Carrillo Navarro: Writing -- review \& editing, Writing -- original draft, Visualization, Validation, Software, Project administration, Methodology, Investigation, Formal analysis, Data curation, Conceptualization. Pablo Garc\'ia-Mart\'in: Writing -- review \& editing, Supervision, Project administration, Methodology. Ren\'e Duffard: Writing -- review \& editing, Supervision, Resources, Project administration, Funding acquisition. Juan Luis Orellana Montes de Oca: Writing -- review \& editing, Software, Methodology, Investigation. Oscar Ortega R\'ios: Writing -- review \& editing, Software.

\section*{Code, data, and materials availability}
This work builds upon standard open-source software libraries and publicly available pre-trained machine learning frameworks. Requests for source code access may be submitted to the corresponding author and will be considered on a case-by-case basis, subject to institutional approval.

\section*{Declaration of generative AI and AI-assisted technologies in the manuscript preparation process}
During the preparation of this work, the authors used ChatGPT (OpenAI) in order to assist with language refinement, text structuring, and clarification of technical explanations. After using this tool, the authors reviewed and edited the content as needed and take full responsibility for the content of the published article.

\section*{Declaration of competing interest}
The authors declare that they have no known competing financial interests or personal relationships that could have appeared to influence the work reported in this paper.

\section*{Data availability}
Data will be made available on request.


\bibliographystyle{cas-model2-names}

\nolinenumbers
\bibliography{cas-refs}

\end{document}